\documentclass[useAMS,usenatbib]{mn2e}
\usepackage{amsmath}
\usepackage{amssymb}
\usepackage{times}
\usepackage{graphicx}
\usepackage{wasysym}
\usepackage{mathtools}

\title[Dark Matter Annihilation in the Circumgalactic Medium]{Dark Matter Annihilation in the Circumgalactic Medium at High Redshifts}

\author[Sch\"{o}n et. al.]{S. Sch\"{o}n$^{1}$\thanks{E-mail:
     sschon@student.unimelb.edu.au},  K. J. Mack$^{1, 2, 3}$, J. S. B. Wyithe$^{1, 2}$\\ \\ $^1$ School of Physics, The University of Melbourne, Parkville, VIC 3010, Australia \\ $^2$ ARC Centre of Excellence for All-Sky Astrophysics (CAASTRO) \\ $^3$ ARC Centre of Excellence for Particle Physics at the Terascale (CoEPP) \\}

\begin{document}

\maketitle

\begin{abstract}
Annihilating dark matter (DM) models offer promising avenues for future DM detection, in particular via modification of astrophysical signals. However when modelling such potential signals at high redshift the emergence of both dark matter and baryonic structure, as well as the complexities of the energy transfer process, need to be taken into account. In the following paper we present a detailed energy deposition code and use this to examine the energy transfer efficiency of annihilating dark matter at high redshift, including the effects on baryonic structure. We employ the PYTHIA code to model neutralino-like DM candidates and their subsequent annihilation products for a range of masses and annihilation channels. We also compare different density profiles and mass-concentration relations for $10^5-10^7\mathrm{M}_{\odot}$ halos at redshifts $20$ and $40$. For these DM halo and particle models, we show radially dependent ionisation and heating curves and compare the deposited energy to the halos' gravitational binding energy. We use the ``filtered'' annihilation spectra escaping the halo to calculate the heating of the circumgalactic medium and show that the mass of the minimal star forming object is increased by a factor of 2-3 at redshift $20$ and 4-5 at redshift $40$ for some DM models. 
\end{abstract}

\begin{keywords}
Cosmology -- Particle Physics, Dark Matter, Structure Formation 
\end{keywords}

\section{Introduction}

The potential impact of annihilating and decaying dark matter (DM) in the context of standard astrophysics has shown itself to be a promising avenue in the exploration of dark matter phenomenology \citep{Chen2004, Padmanabhan2005, Furlanetto2006, Slatyer2013, Valdes2013, Bertone2016}. This is in part because annihilating models are well motivated from a particle physics perspective, often arising naturally within new theoretical frameworks for physics beyond the Standard Model  \citep{Bergstrom2012, Roos2012, Bertone2005}, but also because their non-gravitational interactions make them attractive candidates for possible future detection.

Indirect searches for annihilating and decaying dark matter focus on the detection of the highly energetic annihilation products which maybe used to infer the existence of the parent particles. As any non-gravitational interaction such as annihilation is assumed to be at or around the weak scale, in general a high density region, such as found at the centre of Milky Way, is required for detectable particle excesses \citep{Prada2004, Grasso2009, Slatyer2012}.

While the Galactic Centre presents a relatively accessible over-dense region of dark matter, it is also home of one of the most astrophysically rich regions in the Milky Way. This highlights an ongoing complication with these kind of experiments as they rely on meticulous treatment of the underlying astrophysics in order to disentangle a potential dark matter signal. More often that underlying physics is not yet fully understood or the signal may even be degenerate with a standard process. Similarly, uncertainty in the underlying dark matter structure \citep{Mack2014} makes for further complications, though non-detections from dwarf spheroidal galaxies occupying the Milky Way's sub-haloes have been used to constrain the annihilation cross-section for some dark matter models.

\begin{table*}
\centering
\begin{tabular}{l  c c c c}
\hline
DM Model & $m_{dm}$ &   Annihilation  & Cross-section \\
                 & [$\mathrm{GeV}$] &  Channel  &  [$\mathrm{cm^{3}/\mathrm{s}}$]  \\
\hline
DM1 & $0.13 - 100$ &  $e^{-}$ & $10^{-28} - 7.7 \times 10^{-26}$\\
DM2 & $5, 20, 50$  & $\tau$ & $3.8 \times 10^{-27}$, $1.5 \times 10^{-26}$, $3.8 \times 10^{-26}$ \\
DM3 & $5, 20, 50$ &  $\mu$ & $3.8 \times 10^{-27}$, $1.5 \times 10^{-26}$, $3.8 \times 10^{-26}$ \\
DM4 & $5.5, 20, 50$ & $q$ & $4.2 \times 10^{-27}$, $1.5 \times 10^{-26}$, $3.8 \times 10^{-26}$\\
DM5 & $83, 110$ & $W$ &  $6.4 \times 10^{-26}$, $8.5 \times 10^{-26}$\\
\hline 

\end{tabular}
\caption{Summary of the different dark matter annihilation models considered in this paper.}
\end{table*}

Instead of focusing on a single, dense, and sufficiently near region of dark matter to detect an excess in the associated annihilation products, another possibility is to look for the potential global impact dark matter annihilation may have on standard cosmic evolution. One of the most exciting is the potential modification of the 21cm signal associated with the Epoch of Reionsation. There are a number of ways in which the additional injected energy from DM annihilation and decay could produce changes including direct heating and ionisation of the intergalactic medium (IGM) \citep{Evoli2014,  Poulin2015, Lopez2016} as well as secondary effects from DM modification of baryonic structure. A number of next-generation radio telescopes such as the Square Kilometre Array (SKA) \citep{SKA2013}, Precision Array for Probing the Epoch of Reionisation (PAPER) \citep{PAPER2013} and the Low-Frequency Array (LOFAR) \citep{LOFAR2012} are preparing to probe the Epoch of Reionisation using the 21-cm signal \citep{Mesinger2014}.

The following work is motivated by the prospect of using global signals such as the 21cm line of neutral hydrogen or other observables such as galaxy properties to constrain annihilating dark matter models. One of the pivotal aspects of this is the integration of dark matter annihilation with standard astrophysical processes, such as star formation and the heating of the intergalactic medium, as well as potential complex feedback mechanisms between forming structure and the injected annihilation products. The impact of DM annihilation on individual haloes has previously been discussed in \citet{Ripamonti2007, Spolyar2008, Natarajan2009, Gondolo2013} as well as our previous work \citep{Schon2015} (from here on S15). 

In S15 we performed a first order calculation to gauge the impact of dark matter annihilation on early structure formation. We employed a simplified energy transfer process based on the MEDEAII code \citep{Evoli2012} to cover a wide parameter space and a range of DM halo and particle models. We showed that in small, high redshift objects the energy input from dark matter annihilation over the Hubble time is comparable to the gravitational binding energy of the object. In this paper we foremost look to update the energy transfer code used in S15 to allow the full evolution of the injected annihilation products, including the secondary particle cascades. The new version of the code also determines the radial distribution of the deposited energy in and outside the halo.

To demonstrate the use of the code, we revisited the halo parameter space from our previous work. This paper is structured in the following manner: \S 2 and \S 3 outline the dark matter halo and particle models employed throughout the work and \S 4 introduces the Monte Carlo code used to calculate the energy transfer in and around the chosen haloes. The ionisation and heating curves are calculated, showing the radial distribution of deposited energy within the halo and presented in \S 5 along with the comparison to the haloes' gravitational binding energy. Section 6  shows how the initial injection spectra of the different annihilation models is changed by the gas component of the halo and how the energy that escapes into the circumgalactic medium (CGM) can heat the surrounding gas sufficiently to increase the mass of the minimal baryonic object for some haloes. The paper concludes with a discussion in \S 7 of results.

Throughout the paper we take the cosmological parameters to be in keeping with the \citet{Planck} so that $h = 0.71$, $\Omega_{\Lambda,0} = 0.6825$,  $\Omega_{b, 0}h^{2}=0.022068$ and $\Omega_{dm, 0}h^{2}=0.12029$.

\section{Dark Matter Model}
We employ the same effective, self-annihilating, neutralino-like DM particle model as in S15, albeit with an additional $130$ MeV, $10$ GeV and $100$ GeV particle annihilating directly to an electron/positron pair. A natural choice of cross-section is close to that of the weak interaction as this gives the correct current day density if dark matter is assumed to be a thermal relic. In this work it is assumed that dark matter is cold and the velocity averaged annihilation cross-section is taken as constant, regardless of environment. We set the annihilation cross-section for the $130$ MeV model annihilating to an electron/positron pair to be $1.0 \times 10^{-28}$ $\mathrm{cm}^{-3}\mathrm{s}^{-1}$ to be in keeping with the latest dark matter annihilation limits from CMB measurements \citep{Madhavacheril2014, Planck2016}. The cross-section for all other mass models is adjusted so that the total annihilation power is the same as that of the $130$ MeV model to allow for comparison of the energy transfer efficiency of DM models with different masses and annihilation channels.

\subsubsection{Annihilation Power}
The annihilation power per unit volume of the dark matter component is then given by:
\begin{eqnarray}
P_{dm}(\mathbf{x}) = \frac{c^{2}}{m_{dm}}  \langle v \sigma  \rangle \rho_{dm}^{2}(\mathbf{x}).
\end{eqnarray}
where $m_{dm}$ is the mass of the DM particle, $\rho_{dm}$ is the DM volume density at point $\mathbf{x}$ in the halo and $\langle v\sigma \rangle$  the velocity averaged annihilation cross-section.
\subsubsection{Annihilation Channels}
We consider annihilation via muons, quarks, tau lepton and W bosons as well as annihilation straight to electron positron pairs. In the case of unstable annihilation products, the particles
 were further evolved in PYTHIA \citep{PYTHIA2006, PYTHIA2008} where we use a $e^{-}, e^{+}$ proxy, until only stable particles remain (see Appendix A of S15 for detailed final electron, positron and photon energy distributions). The different DM annihilation models are summarised in Table 1. 

The annihilation process also produces a significant fraction of neutrinos. However, as these only interact on the weak scale with the surrounding gas, we assume they do not contribute in a significant fashion to the energy transfer process and we thus do not consider them further in this work.

\section{Halo Models}
In order investigate the degree to which energy from DM annihilation transfers energy into the halo and CGM, we require models for the DM  and gas distribution within the halo as well as a density profile for the gas immediately surrounding the halo. We note that since we are primarily interested in high redshift haloes with masses well below the general resolution limit of N-body simulations there is a considerable degree of uncertainty involved in these models \citep{Mack2014}.

\begin{figure}
\centering
\includegraphics[scale=0.5]{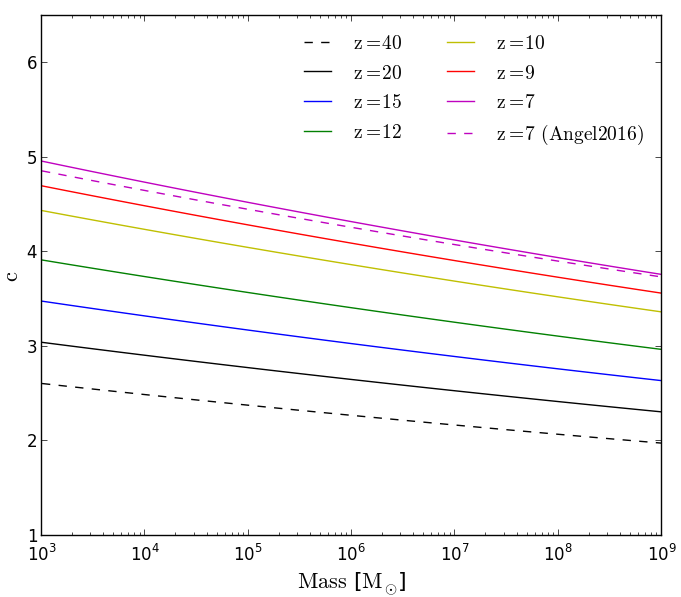}
\caption{The fiducial mass-concentration relation used in this work. Solid lines show the semi-analytic model from \citet{Correa2015} at various redshifts with the dashed black curve showing an extrapolation to redshift 40. The dashed pink line shows results from \citet{Angel2015} which was extracted from N-body simulations.}
\label{Concentration}
\end{figure}

\subsection{DM Density Profiles}
The radially dependent density profile of DM haloes has been shown to be well described by a universal profile wholely determined by the mass and redshift of the halo. This profile was first fit as a broken power law by \citet{Navarro1995} and concequently named the \textquoteleft Navarro-Frenk-White\textquoteright profile,
\begin{eqnarray}
\rho_{NFW} = \frac{\rho_{NFW,0}}{\frac{r}{r_{s}}(1 + \frac{r}{r_{s}})^{2}}.
\end{eqnarray}
Later work has shown that this fit can further be improved \citep{Merritt2005, Merritt2006}  by the Einasto profile \citep{Einasto1965}
\begin{eqnarray}
 \rho_{E} = \rho_{E,0}e^{-\frac{2}{\alpha_{e}}[(\frac{r}{r_{s}})^{\alpha}-1]}.
\end{eqnarray}
Both the NFW and Einasto profile fit haloes with high density cusps. For comparison we also include the Burkert profile \citep{Burkert1995} which has a flat core inwards of the scale-radius $r_{s}$.
\begin{eqnarray}
 \rho_{B} = \frac{\rho_{B,0}}{(1+\frac{r}{r_{s}})(1+\frac{r^{2}}{r_{s}^{2}})}
\end{eqnarray} 
 where $\rho_{NFW,0}$, $\rho_{E,0}$ and $\rho_{B,0}$ are the appropriate dark matter density normalisation parameters for each profile. For all models,
\begin{eqnarray}
r_{s} = \frac{r_{vir}}{c}
\end{eqnarray}
Here $r_{vir}$ is the virial radius, which is defined as enclosing 200 times the critical density $\rho_{crit}$. The Einasto profile contains an addition shape parameter  $\alpha_{e} = 0.17$ which is in general only weakly dependent on the mass and redshift of the halo and taken to be a constant. 

\subsubsection{Mass-concentration relation}
The mass concentration parameter $c$ is a free parameter of the profiles described above, which sets the turn-over of the density gradient. It's of particular interest in dark matter annihilation calculations because it can significantly enhance the overall power output in individual haloes by increasing the DM density at its core. While the concentration parameter is well described \citep{Comerford2007, Duffy2008, Ludlow2013} for galaxy hosting haloes at low redshifts, it is less clear how the parameter behaves in the mass-redshift parameter space reviewed in this work. Simply extrapolating these mass-concentration relations to the required masses and redshifts can lead to haloes with extremely dense cores, which in turn has a significant impact on the DM annihilation signal. 

We use the semi-analytic model based mass-concentration relation of \citet{Correa2015} which predicts $c$ for redshifts $z = 0-20$  and masses $\mathrm{M}/\mathrm{M}_{\odot} = [10^{-2}, 10^{16}]$:
\begin{eqnarray}
\mathrm{log}_{10}c &=& \alpha + \beta\mathrm{log}_{10}(\mathrm{M}/\mathrm{M}_{\odot})
\end{eqnarray}
\begin{eqnarray*}
\alpha &=&  1.226 - 0.1009(1 + z) + 0.00378(1 + z)^{2} \\
\beta  &=&  0.008634 - 0.08814(1 + z)^{-0.58816} .
\end{eqnarray*}
Figure 1 shows the mass-concentration relation produced by \citet{Correa2015} at different redshifts (solid coloured lines). In addition the mass-concentration relation at redshift $7$ \citep{Angel2015} from N-body simulations is shown in comparison by the dashed pink line. The dashed black line shows $c$ extrapolated from the semi-analytic model to redshift $40$.  

Lastly we summarise the different halo models in Table 2. Throughout the paper the ``fiducial'' and ``Burkert'' models refer to haloes with the Correa mass-concentration relation and an Einasto and Burkert density profile respectively. The ``high'' and ``Burkert-high'' models refer to haloes with Einasto and Burkert profiles as well as an increased $c$ parameter, here taken to be double that of the Correa model.

\subsubsection{Dynamic Haloes}
At this point a brief mention should be made about the assumption that the haloes under consideration here are in a static state of equilibrium, and that both the underlying dark matter structure and the baryonic content of the halo remain unchanged. More realistically, haloes after collapse will grow through a combination of merger events and inflowing gas. At high redshifts especially, the rapid merger rate means that a significant fraction of haloes are not relaxed, and moreover, recovery takes place over a number of dynamical times \citep{Poole2016}.

The disruption of both the dark matter and gas density distributions associated with mergers could subsequently impact both the dark matter annihilation power and the absorption efficiency. In addition self-consistent modelling of the halo incorporating the potential ongoing effects due to the halo self-heating, such as expulsion of gas, is not accounted for. These concerns are anticipated to be included in future work.

\subsection{Gas Model}

\subsubsection{Baryonic Density Distribution within the Halo}
In S15, the gas component of the halo was assumed to follow that of the dark matter distribution. We update our gas profile to that of a pseudo-isothermal sphere, as it makes a reasonable first order approximation for the mass distribution of halo-type objects \citep{Binney1987} and avoids a divergent core density in the case of the NFW profile.
\begin{equation}
\rho_{gas}(r)  = \frac{\rho_{bar,0}}{r_{0} +r^{2}}.
\end{equation}
Here $\rho_{bar,0}$ gives the appropriate baryon density normalisation and $r_{0} = r_{vir}/10^4$. The total baryon fraction of the halo is taken to be $f_{b} = 0.15$ for all haloes so that $M_{gas}  =  \frac{f_{b}}{1 - f_{b}}M_{DM}$, where $M_{gas}$ and $M_{DM}$ are the total gas and DM mass contained in the halo.
\subsubsection{Circumgalactic Density Distribution}
The gas density of the CGM gas follows the fit by \citet{Dijkstra2007, Bruns2012}
\begin{eqnarray}
\rho_{CGM}(r, r_{vir}) = \begin{cases} 20 \bar{\rho} (r/r_{vir})^{-1} \hfill &  r < 10r_{vir}, \\
                                                  \bar{\rho}[1 + \text{exp}(2-r/5r_{vir})] \hfill & r \geq 10r_{vir} .\\
                      \end{cases}
\end{eqnarray}
where $\bar{\rho}$ is the average gas density of the IGM. We consider haloes at redshifts corresponding to the very onset of star and galaxy formation and assume that the gas in and around the haloes can be treated as neutral and only consisting of Hydrogen and Helium. As the gas begins to be ionised, the energy deposition would be pushed towards heating as Coulomb interactions become prominent while electro- and photo-ionisation are suppressed. 

\begin{table}
\centering
\begin{tabular}{l  c c c c}
\hline
Halo Model & mass &   redshift  &  c & profile  \\
                 & $\mathrm{log}_{10}[\mathrm{M}_{\odot}$] & [$\mathrm{z}$ ] &   &     \\
\hline
Fiducial & $5-7$ & $20, 40$ & Correa & Einasto \\
Burkert & $5-7$ & $20, 40$ & Correa & Burkert\\
High & $5-7$ & $20, 40$ & Correa $\times 2$ & Einasto\\
Burkert-high & $5-7$ & $20, 40$ &Correa $\times 2$  & Burkert\\
\hline 
\end{tabular}
\caption{Summary of the different DM halo models considered in this work.}
\end{table}

\section{Energy Transfer Code }
We update the energy transfer code used in S15 to more fully model the paths of the injected particles (as well as the secondary particles created in the subsequent cascades) throughout the halo. In addition, the updated code allows us to record the spatial distribution of the deposited energy in the form of heating, ionisation and Lyman photons as well as keep a record of the energy distribution of the particles that escape the halo.

\subsection{Processes}
In our treatment of the relevant interaction processes we closely follow the treatment of \citet{Evoli2012} and \citet{Slatyer2009}. 

\begin{figure}
\centering
\includegraphics[scale=0.45]{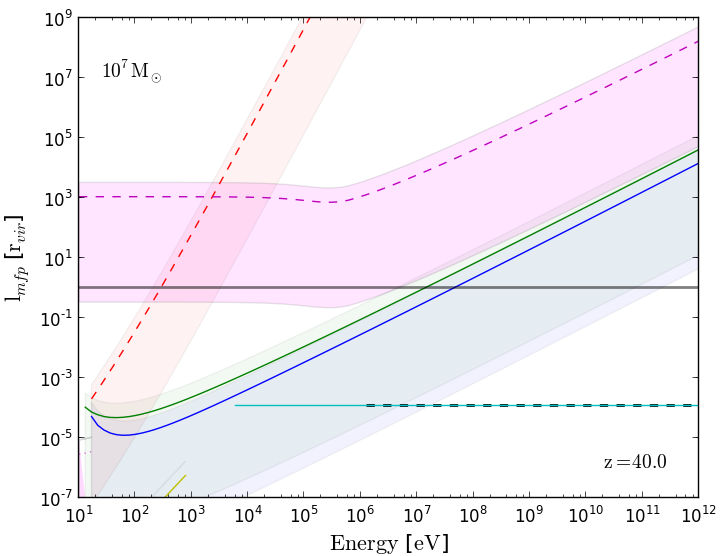}
\caption{Mean free paths in units of the virial radius for the various interactions undergone by injected particles within a neutral gas $10^7\mathrm{M}_{\odot}$ halo at redshift $40$. Solid lines show interactions for charged particles and dashed for photons. The grey line indicates the virial radius of the halo. Processes shown are photo-ionisation (red), Compton scattering (magenta) and pair creation off of the CMB (black) for photons and ionisation and excitation (green and blue) for electrons, as well as inverse Compton scattering (cyan) and annihilation via Positronium for positrons (purple). In all cases, lines show the mean free path for the process at the average density of the halo, while the shaded regions either side indicate the mfp at $\mathrm{r}_{vir}$ and $\mathrm{r}_{vir}$/100. For reference the yellow curve shows the mean free path for Coulomb scattering in an equivalent, fully ionised halo.}
\label{Concentration}
\end{figure}

\subsubsection{Electrons and Positrons}
At high energies, electrons and positrons predominantly lose energy through inverse Compton (IC) scattering off cosmic microwave background (CMB) photons \citep{Blumenthal} while lower energy electrons undergo collisional interaction with the gas atoms. These collisions lead to ionization \citep{Kim1994, Shah1987, Shah1988, Arnaud1985}, atomic excitation \citep{Hirata2006, Chuzhoy2007, Stone2002, Bransden1976, Fisher1997}, and Coulomb scattering \citep{Shull1985, Furlanetto2010, Spitzer1969}. Recombination \citep{Karzas1961} is also included, though subdominant compared to the other processes.

At high energies positrons behave in the same manner as electrons. At low energies they also undergo collisional interactions in addition to forming positronium via charge exchange \citep{Guessoum2005, Heitler1954}.  

\subsubsection{Photons} 
In a similar fashion photons photo-ionise the surrounding gas \citep{Verner1996}, Compton scatter off free electrons \citep{Heitler1954, Chen2004} and at high enough energies undergo electron-positron pair-production (PP). We here distinguish the cross-section for PP off of the CMB \citep{Agaronyan1983, Ferrigno2005}, atomic and ionised Hydrogen and Helium, as well as free electrons \citep{Zdziarski1989, Motz1969, Joseph1958}. Photons also undergo photon-photon scattering where the injected photon transfers part of its energy to a CMB photon \citep{Svensson1990}, however this process is highly sub-dominant in the redshift range considered here. 

The mean free path (mfp) at mean gas density for selected processes undergone by injected electrons, positrons and photons are outlined in Figure 2 for a neutral $10^7\mathrm{M}_{\odot}$ halo at redshift $40$. Solid lines show interactions for the leptons while the dashed show those for photons. The dashed red, magenta and black lines show photo-ionisation, Compton scattering and pair creation off of the CMB photon field respectively. Electro-ionisation and -excitation are shown in green and blue, IC scattering in black and annihilation via positronium for positron in purple. For a fully ionised halo, the yellow line shows the mean free path for Coulomb interaction. For each process, the lines indicate the mfp at the average density of the halo while the shading either side of the curves shows the mean free path at $r_{vir}$ and $r_{vir/100}$.

The grey line is indicative of the halo's virial radius. Particles with energies such that their interaction mean free paths are no longer smaller or comparable to the virial radius will readily escape the halo without either interacting with the gas and/or depositing energy. For the DM annihilation products considered here, the primary interaction of the injected electrons and positrons is IC scattering (injected photons first undergo electron/positron pair-creation off of the CMB photon field). The actual energy transfer process occurs via the secondary particle produced in those interactions, in this case, the up-scattered IC photons. Therefore the energy spectrum of the secondary cascade products induced by the injected particles plays an important role in determining the relative efficiency with which the different dark matter models are able to deposit energy into the halo.

\subsection{Particle Evolution}
The step-wise evolution of the particle follows the methodology of the GEANT4 simulation toolkit \citep{Agostinelli2003, Allison2006}. At the beginning of each calculation the code takes a particle (in this case an electron, photon or positron) with a given energy and injects it at point $\mathbf{x}$ in direction ($\theta, \phi$). The particle then takes a straight line step of size determined by the step-size function in this specified direction. The code then selects a random point along this straight line step and uses the densities at this point to calculate the relevant mean free paths which characterise the distance the particle travels before said interactions take place. For each potential process random numbers are used to determine at what point the interaction takes place. If none of the interactions take place within the step-size, the particle travels to the end of the straight line, the next step is taken, and the process is repeated. If one or more interactions do occur within the step, the process corresponding to the shortest distance to interaction will be executed. At the end of the process, the code records the location of any energy that has been deposited as well as the type, energy, position and orientation of the remaining/newly created particles. The process repeats until either all particles' energies drop below all interaction thresholds (or if one is defined, the particle escapes a pre-defined spatial boundary). 

\begin{figure}
\centering
\includegraphics[scale=0.45]{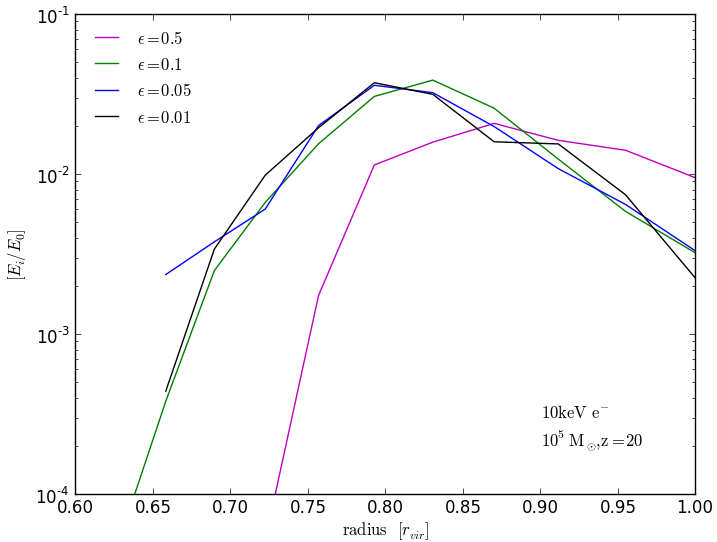}
\caption{Outputs for a $10$ keV electron injected into $10^5\mathrm{M}_{\odot}$ halo at redshift $20$ for different step size scale-parameters $\epsilon$. Magenta, green, blue and black curves respectively refer to $\epsilon=0.5$, $0.1$, $0.05$ and $0.01$. Results are given as the fraction of the original injected particle energy deposited in each respective radial bin.}
\end{figure}

The initial particle can be evolved as often as required to achieve the necessary level of convergence. At the end of the Monte Carlo run the code outputs the average energy deposited into heating, ionisation and Lyman-alpha photons in each pre-defined spatial bin as a fraction of the original injected particle's energy. If a physical boundary is defined the code will give a similarly normalised spectrum of the escaped electrons, positrons and photons. 

In this calculation we assume both the halo and the CGM are spherically symmetric. We thus define the stepsize function the following way: 
\begin{eqnarray}
S(\mathbf{r}) = \epsilon_{0} \mathbf{r} 
\end{eqnarray}
where $\epsilon_{0}$ is an adjustable parameter setting the length of the step. Figure 3 shows the code output for a $10^4$ keV electron in a $10^5\mathrm{M}_{\odot}$ halo at redshift $20$ for different values of $\epsilon$. Results are given as the fraction of the original injected particle energy deposited in each respectively radial bin. The magenta, green, blue and black curves show results for $\epsilon = 0.5$, $0.1$, $0.05$ and $0.01$ respectively. For $\epsilon < 0.1$, the results appear to converge. In contrast, for larger $\epsilon$ the change in density across the resultant step-size is no longer sufficiently close to constant. For a radially, monotonically decreasing density the code will then under-estimate the mean-free paths for particles heading towards the centre of the halo, while over-estimating those for particles heading outwards resulting in the shifted, magenta curve shown in Figure 3. An $\epsilon=0.01$ was used throughout this paper.

\section{Halo Self-heating}
We now use the code to calculate the energy deposited within $10^5$, $10^6$ and $10^7\mathrm{M}_{\odot}$ haloes at redshifts $20$ and $40$.
\subsection{Heating and Ionisation}
In order to determine the total energy transfer for each DM model, the deposition fractions for each particle species $\alpha$, with energy $E_{j}$ injected at radius $r_{i}$ are first calculated, 
\begin{equation} 
\begin{split}
\mathbf{h}_{ijk}^{\alpha}, \quad \mathbf{x}_{ijk}^{\alpha}, \quad \mathbf{l}_{ijk}^{\alpha}\\
\mathbf{e}_{ijk}^{\alpha}, \quad \mathbf{f}_{ijk}^{\alpha}, \quad \mathbf{p}_{ijk}^{\alpha}
\end{split}
\end{equation}
Here $\mathbf{h}$, $\mathbf{x}$ and $\mathbf{l}$ denote the fraction deposited as heating, ionisation and Lyman photons at each radius $r_{k}$, and $\mathbf{e}$, $\mathbf{f}$ and $\mathbf{p}$ the energy distribution of escaped electrons, photons and positrons in energy bin $E_{k}$. These act as elements of a kind of pseudo basis which are then weighted by the appropriate halo and DM models. The total, halo-summed energy deposition for a particle $\alpha$ with energy $E_{j}$ is then given by,
\begin{equation}
\mathbf{H}_{jk}^{\alpha} = \sum_{i} w_{i} \mathbf{h}_{ijk}^{\alpha}, \quad \mathbf{\chi}_{jk}^{\alpha} = \sum_{i} w_{i} \mathbf{x}_{ijk}^{\alpha}, \quad \mathbf{L}_{jk}^{\alpha} = \sum_{i} w_{i} \mathbf{l}_{ijk}^{\alpha},
\end{equation}
where $w_{i}$ is the fraction of the total dark matter annihilation power produced within the shell at radius $r_{i}$. The coefficients $w_{i}$ will naturally vary depending on the halo model, with concentrated models (such as those with an Einasto profile) injecting more of their total annihilation energy close to the centre of the halo while in haloes with the flat-core Burkert profile the injected energy is distributed closer to the edge. This means that more concentrated haloes not only have greater total annihilation power, they are also more efficient at energy transfer as more of their energy is injected close to the haloes' centre where the gas density is greatest.  

The code outputs can now be used to calculate the radially dependent heating $\mathfrak{H}$, ionisation $\mathfrak{X}$ and Lyman photon $\mathfrak{L}$ production for different dark matter models,
\begin{equation}
\begin{split}
\mathfrak{H}_{k} = \sum_{\alpha} f_{j\alpha} \sum_{j} \mathbf{H}_{jk}^{\alpha}\\
\mathfrak{X}_{k} = \sum_{\alpha}  f_{j \alpha}\sum_{j} \mathbf{\chi}_{jk}^{\alpha}\\
\mathfrak{L}_{k} = \sum_{\alpha} f_{j  \alpha} \sum_{j} \mathbf{L}_{jk}^{\alpha},
\end{split}
\end{equation} 
where $f_{j\alpha }$ is the fraction of the annihilation spectrum carried by particle $\alpha$ with energy $E_{j}$. $f_{j\alpha}$ depends on the annihilation channel of the dark matter model.

\begin{figure*}
\centering
\includegraphics[scale=0.6]{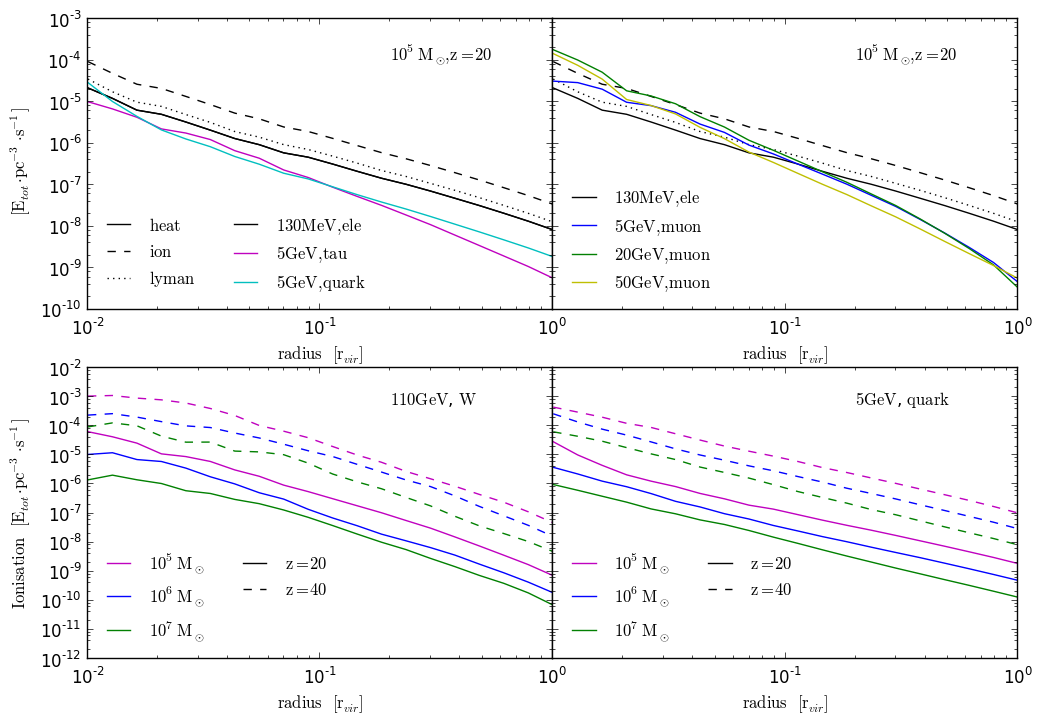}
\caption{Heating and ionisation curves for different haloes and dark matter models. All results are given as the fraction of the total annihilation power of the halo deposited per unit volume. Upper plots show the outputs for a $10^5\mathrm{M}_{\odot}$ halo at redshift 20. The figure on the left shows a $5$ GeV particle annihilating via tau leptons (magenta) and quarks (cyan) while the figure on the right shows models annihilating via muons with $5$ GeV, $20$ GeV and $50$ GeV shown in blue, green and yellow. In both plots the black curves show the $130$ MeV model annihilating directly to an electron/positron pair with the solid, dashed and dotted lines showing energy going into heating, ionisation and Lyman photons for reference. The lower, left plots shows results for a $110$ GeV DM particle annihilating via W bosons and the lower right shows those for a $5$ GeV particle annihilating via quarks, In each of the lower plots, magenta, blue and green curves show the ionisation curves for $10^5$, $10^6$ and $10^7\mathrm{M}_{\odot}$ haloes where solid lines indicate outputs at redshift $20$ and dashed those at redshift $40$.}
\end{figure*}

Figure 4 shows selected heating and ionisation curves for different halo and dark matter particle models. All results are given as the fraction of the total annihilation power of the halo deposited per unit volume. Upper plots show the outputs for a $10^5\mathrm{M}_{\odot}$ halo at redshift 20. The figure on the left shows a $5$ GeV particle annihilating via tau leptons (magenta) and quarks (cyan) while the figure on the right shows models annihilating via muons with $5$ GeV, $20$ GeV and $50$ GeV shown in blue, green and yellow. In both plots the black curves shows the $130$ MeV model annihilating directly to an electron/positron pair with the solid, dashed and dotted lines showing energy going into heating, ionisation and Lyman photons for reference. 

As mentioned previously the relative energy transfer efficiency of the various dark matter models is strongly dependent on the energy distribution of the secondary cascade particles. Of particular note is the $130$ MeV model annihilating via an electron/positron pair shown in black in the upper row of Figure 4. Electrons/positrons injected at $130$ MeV will produce up-scattered IC photons with energies low enough to undergo photo-ionisation so that even the lower density regions at the edge of the halo can be heated. This, along with the fact that none of the annihilation power is lost to neutrinos, makes it a very attractive DM candidate as far as possible detection goes.

For somewhat heavier DM models ($1-5$ GeV) the peak of up-scattered IC photons will fall into the energy regime of Compton scattering which is only efficient at the high gas densities found at the centre of the halo. Even heavier candidates will eventually produce IC photons sufficiently energetic to undergo pair-creation themselves leading to more complex cascades. In both of those scenarios the energy deposition in the outer parts of the haloes is less efficient due to a significant fraction of the resultant cascade particles falling into energy regimes in which interaction with the gas is poor. Lastly for neutral gas, the majority of the deposited energy is partitioned into ionisation as is indicated by the black dashed line in the upper plots of Figure 4. However as the halo becomes ionised, Coulomb heating very rapidly becomes a dominant interaction process and the energy partition becomes skewed towards heating instead.

The lower left plot shows results for a $110$ GeV DM particle annihilating via W bosons and the lower right shows those for a $5$ GeV particle annihilating via quarks, In each of the lower plots, magenta, blue and green curves show the ionisation curves for $10^5$, $10^6$ and $10^7\mathrm{M}_{\odot}$ haloes where solid lines indicate outputs at redshift $20$ and dashed those at redshift $40$. As expected the energy deposition efficiency at higher redshifts is greater due to the increase in both the gas and CMB number density. 

\subsubsection{Lyman Photons}
As shown in Figure 4, our code also tracks the production of Lyman photons. Photons with energy below the Lyman limit (13.6eV) are assumed to cease direct interaction with their surrounding medium. However their production through dark matter annihilation can still play an important role in baryonic structure formation when taking gas cooling into account. For instance photons with energies between 11.2 and 13.6eV are classified as Lyman-Werner photons. These photons maybe absorbed by molecular hydrogen and through excitation and subsequent radiative decay lead to its disassociation. Before the production of metals, the formation and destruction of $\mathrm{H}_{2}$ plays an important role in molecular cooling at high redshifts (Abel et al., 1997) and are thus vital for early star (Abel et al., 2002) and galaxy formation (Bromm et al., 2009).

\subsection{Binding Energy Comparison}
The above results allow for the comparison of the halo's gravitational binding energy and the DM energy input to be revisited. However, unlike in the work conducted in S15, the comparison will be made for individual shells rather than the total halo. The energy produced due to DM annihilation over the Hubble time $t_{H}$, in the spherical shell between $r_{k}$ and $r_{k+1}$ is
\begin{equation}
U_{dm,shell}(r_{k}) = \int P_{dm}(r_{k}) \mathrm{d}S \cdot t_{H}.
\end{equation}
The gravitational binding energy is given by,
\begin{equation}
U_{G,shell}(r_{k}) = \frac{G M_{shell}(r_{k}) M(<r)}{r_{k}}
\end{equation}
so that the ratio of the energy going into heating and the gravitational binding energy is
\begin{equation}
R_{g}(r_{k}) = \frac{U_{dm,shell}(r_{k})}{U_{G, shell}(r_{k})} \cdot \mathfrak{H}_{k}
\end{equation}
where $\mathfrak{H}$ is the heating term previously calculated.

\begin{figure*}
\centering
\includegraphics[scale=0.85]{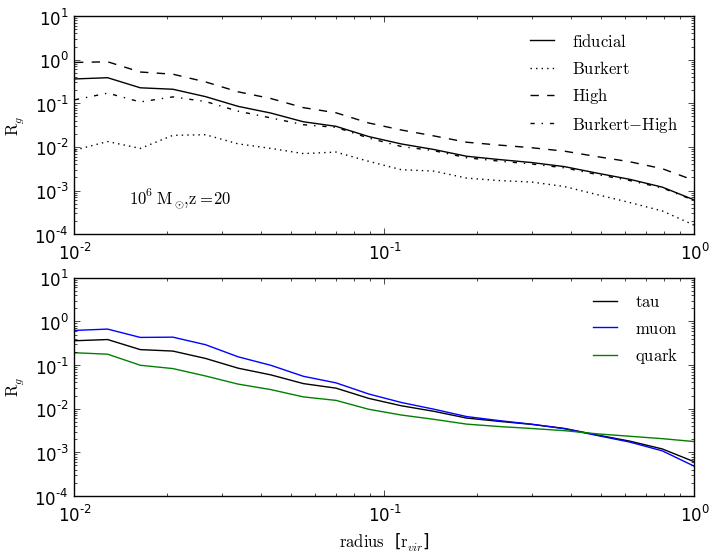}
\caption{The ratio $R_{g}$ comparing the energy deposited as heat over the Hubble time with the gravitational binding energy of shells at different radii in a $10^6\mathrm{M}_{\odot}$ halo at redshift $20$. The upper plot compares different DM haloes models for a $20$ GeV DM particle annihilating via tau leptons. The fiducial halo model is shown by the solid black line and the Burkert model is shown by the dotted curve. The High and High-Burkert models are given by the dashed and dot-dashed curves respectively. The lower plot shows $R_{g}$ for the  fiducial halo model but with the black, blue and green lines showing models annihilating via tau leptons, muons and quarks.}
\end{figure*}

\begin{figure*}
\centering
\includegraphics[scale=0.6]{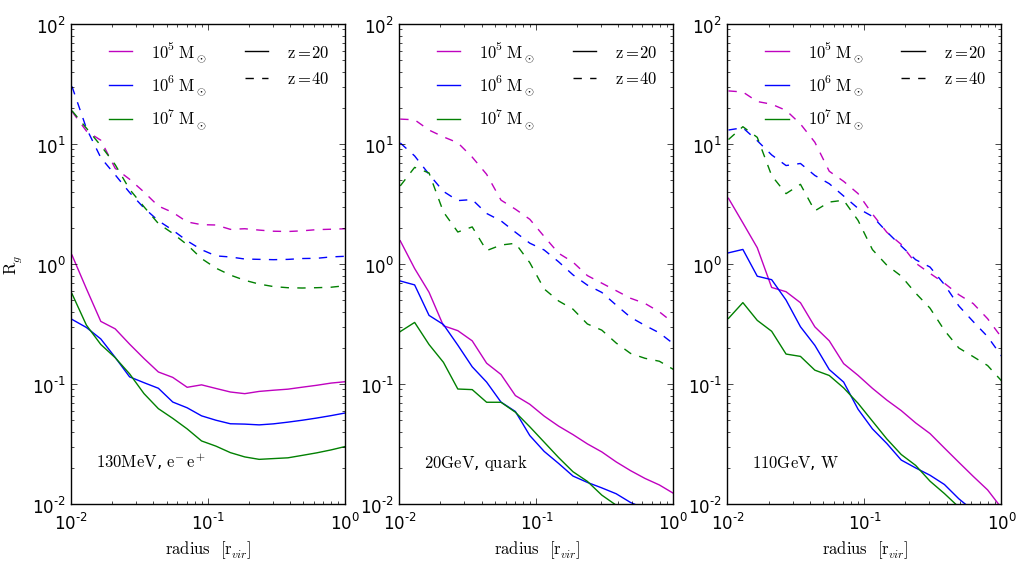}
\caption{The ratio $\mathrm{R}_{g}$ of heating from dark matter annihilation over the Hubble time and gravitational binding energy at different radii. From left to right plots show results for $130$ MeV particle annihilating to an electron/positron pair, $20$ GeV annihilating to quarks and $110$ GeV annihilation to a W boson. As before magenta, blue and green curves correspond to $10^5$, $10^6$ and $10^7\mathrm{M}_{\odot}$ haloes while solid lines refer to results at redshift $20$ and dashed to those at redshift $40$.}
\end{figure*}

Figure 5 shows the comparison $R_{g}$ between the energy deposited over the Hubble time and the gravitational binding energy of the haloes at different radii. For a $10^6\mathrm{M}_{\odot}$ halo at redshift $20$, the upper plots compare different halo models where the solid, dotted dashed and dot-dashed lines show the results for the fiducial, Burkert, High and Burkert-high models respectively. As previously discussed, models with more concentrated mass profiles both produce more annihilation power and are more efficient at depositing energy into the halo. This can be seen here when comparing the High model to the others. Also of interest is that the Burkert-high model is, with exception of the behaviour very close to the halo centre, nigh indistinguishable to the fiducial model. In this scenario doubling the mass-concentration parameter allows the flat-core Burkert profile to mimic the behaviour of the cuspy Einasto profile, again highlighting the potential impact underlying uncertainties in the DM models have on results. The lower plot shows results for different annihilation channels where black, blue and green show the tau lepton, muon and quark models. 

Figure 6 also shows $R_{g}$ but here haloes at different redshift are compared. From left to right plots show results for $130$ MeV particle annihilating to an electron/positron pair, $20$ GeV annihilating to quarks and $110$ GeV annihilation to a W boson. As before magenta, blue and green curves correspond to $10^5$, $10^6$ and $10^7\mathrm{M}_{\odot}$ haloes while solid lines refer to results at redshift $20$ and dashed to those at redshift $40$. Also as before, $R_{g}$ is greater at high redshift due to the higher energy transfer efficiency. The sustained heating of the $130$ MeV model out to the edges of the halo is also demonstrated again.

As a whole this matches the findings of the work in S15, albeit with a more precise estimate of the deposition fractions. \textbf{Having an additional, not insignificant heating source during the early stages of these micro-haloes' development could result in gas being expelled, or alternatively heated to a degree that prevents/delays star formation.} A dynamic model would be required to ascertain how impactful the heating from dark matter turns out to be. While this is beyond the scope of this work, the heating curves calculated here are an important first step towards fully quantifying the impact dark matter annihilation has on early star formation. 

\section{Heating of the Circumgalactic Medium}
While it was here shown that energy comparable to that of the binding energy is deposited inside the halo over the Hubble time, most of the injected energy will escape the halo. In this section the results from this calculation are used to determine the changes to the particle distribution of the original annihilation spectrum as a consequence of traversing the halo. These new spectra are then injected into the circumgalactic medium to establish whether there is sufficient heating to curtail the infall of gas onto the halo, and therefore affect structure formation. 

\citet{Loeb2006} gives an expression for the baryon over-density inside a collapsed dark matter structure after virialisation as 
\begin{equation}
\delta_{b} \equiv \frac{\rho_{b}}{\bar{\rho}_{b}} - 1 = \left( 1 + \frac{6}{5} \frac{T_{vir}}{\bar{T}_{g}}  \right) ^{3/2} -1
\end{equation}
where the background gas temperature $\bar{T}_{g}$ can be written
\begin{equation}
\bar{T}_{gas} \approx 170[(1+z)/100]^{2} \, \mathrm{K}, \quad \text{for  } \mathrm{z}< 160
\end{equation}
Using the result above, a critical value for $\delta_{b}$ can be set to indicate the collapse of the gas. This value is to a certain degree arbitrary so for example when $\delta_{b} > 100$, then $>50$\%  of the gas that would be present if there were no pressure, accrues onto the halo. Having set a critical $\delta_{b}$ to indicate collapse, a minimal baryonic mass can be defined. Regardless of what $\delta_{b}$ is chosen, if the background temperature of the halo $\bar{T}_{g}$ is increased, the virial temperature $T_{vir}$ (which is a proxy for the total mass of the halo) needs to increase along with it to maintain the same $\delta_{b}$. Having a source of heating of the circumgalactic medium such as dark matter annihilation could, if efficient enough, reduce the amount of gas accreted onto the halo or in even more extreme cases raise the minimal halo mass required for baryonic collapse. Unlike for calculations of the Jeans mass, the above expression takes into account shell crossing though it does neglect shock heating which is critical when calculating the DM modified $\delta_{b}$ for the larger of our haloes.

\subsection{Filtered Spectra}
We assume the primary source of heating of the CGM due to dark matter annihilation to come from within the halo itself. When considering the actual energy input into the CGM one has to take into consideration that some of the injected energy will have been lost to the halo's gas component, and some will have been down-scattered through the IC induced particle cascades. The modified annihilation spectra are calculated in much the same manner as the total heating and ionisation curves so that the escaped distribution of electrons, photons and positrons for particle $\alpha$ with energy $E_{i}$, 
\begin{equation}
\mathbf{E}_{jk}^{\alpha} = \sum_{i} w_{i} \mathbf{e}_{ijk}^{\alpha}, \quad \mathbf{F}_{jk}^{\alpha} = \sum_{i} w_{i} \mathbf{f}_{ijk}^{\alpha}, \quad \mathbf{P}_{jk}^{\alpha} = \sum_{i} w_{i} \mathbf{p}_{ijk}^{\alpha} 
\end{equation}
and the total distribution of electrons, photons and positrons for the different annihilation models
\begin{equation}
\begin{split}
\mathfrak{E}_{k} &= \sum_{\alpha} \sum_{j} f_{j \alpha} \mathbf{E}_{jk}^{\alpha}\\
\mathfrak{F}_{k} &= \sum_{\alpha} \sum_{j} f_{j \alpha} \mathbf{F}_{jk}^{\alpha}\\
\mathfrak{P}_{k} &= \sum_{\alpha} \sum_{j} f_{j \alpha} \mathbf{P}_{jk}^{\alpha}
\end{split}
\end{equation}
where $w_{i}$ and $f_{j \alpha}$ are the same coefficients as used in \S 5.1. 

\begin{figure*}
\centering
\includegraphics[scale=0.85]{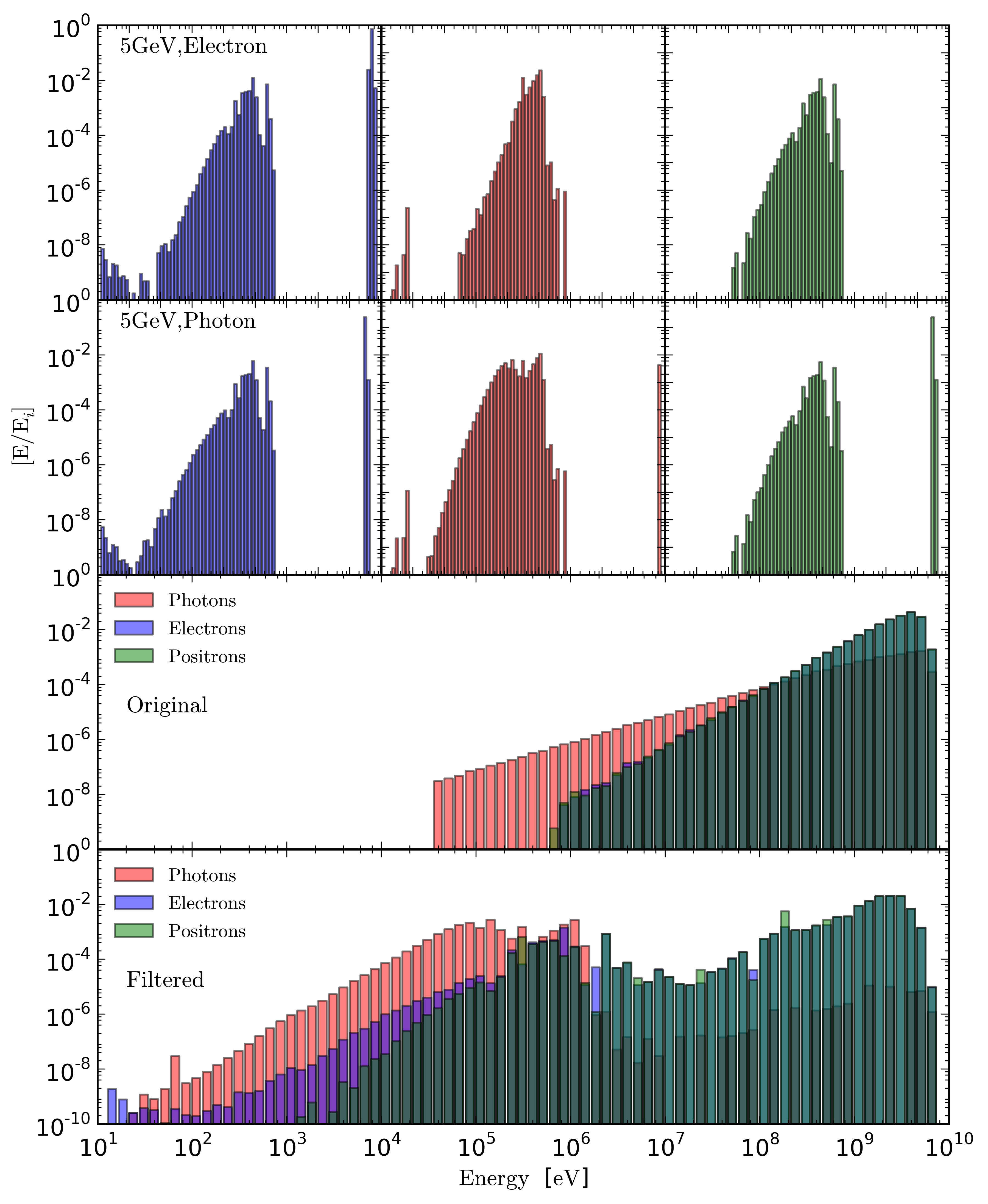}
\caption{Code output showing the energy distribution of particles escaping a $10^6\mathrm{M}_{\odot}$ halo at redshift $20$. The uppermost plot shows the distribution of escaped electrons (left), photons (middle) and positrons (right) for $5$ GeV electrons injected into the halo. The plots directly below show the same but for a $5$ GeV photon. The lower two plots show the total spectrum where the third row shows the original, un-modified injected annihilating spectrum for a $5$ GeV DM particle annihilating via muons. The lowest plot shows the spectrum after the injected particles have passed through the halo. In both plots the pink columns show photons, blue electron and green positrons and results are given as a fraction of the halo's total annihilation power.}
\end{figure*}

Figure 7 demonstrates the code output showing the energy distribution of particles escaping a $10^6\mathrm{M}_{\odot}$ halo at redshift $20$. The upper most plot shows the distribution of escaped electrons (left), photons (middle) and positrons (right) for $5$ GeV electrons injected into the halo. The plots directly below show the same but for a $5$ GeV photon. The lower two plots show the total spectrum where the third row shows the original, un-modified injected annihilating spectrum for a $5$ GeV DM particle annihilating via muons. The lowest plot shows the spectra after the injected particles have passed through the halo. In both plots the pink columns show photons, blue electron and green positrons and results are given as a fraction of the halo's total annihilation power. 

As can be seen in the modified spectra, the original electron/positron distribution is largely retained as only a small fraction of energy is actually lost to the halo, though it is somewhat broadened towards lower energies. In contrast the original high energy photons undergo pair-creation and only those photons injected close to the virial radius are retained. The secondary population of pair-created electrons and positrons can be observed around $10^4-10^6$ eV. Electrons also feature a low energy tail due to ionisation from those IC photons that fall below the pair creation limit. IC scattering will create an extensive secondary population of photons with those IC photons below the pair-creation limit comprising the prominent mid to low energy tail.

While we focus here on the impact on a halo's immediate environment, this modification of the spectrum also has implications for the impact of annihilation products on the mean IGM evolution. The efficiency of energy deposition depends strongly on the energy spectrum of the photons and particles, suggesting that the filtering must be taken into account in studies of DM's impact on reionisation. 

\subsection{Raising Jeans Mass}
The CGM is well approximated by a monatomic, ideal gas for which the internal energy can be written as $U = \frac{3}{2} N k_{b} T$. Here $N$ denotes the total number of particles and $k_{b}$ is the Boltzmann constant. Given the total energy produced through DM annihilation by the halo over the Hubble time, $U_{dm}$ and the heating $\mathbf{H}_{k}$ and ionisation $\mathbf{X}_{k}$ fractions calculated above, the average change in temperature of the shell at $\mathbf{r}_{k}$ is given by,
\begin{equation}
\Delta T(\mathbf{r}_{k}) = \frac{2 \mathbf{H}_{k} U_{dm}}{3 N_{k} k_{b} }.
\end{equation}

To show the potential increase in the minimal baryonic mass we use Equation 16 so that the new $\delta_{b}$ with dark matter annihilation is given by
\begin{equation}
\delta_{mod} = \frac{\rho_{b}}{\bar{\rho}_{b}} - 1 = \left( 1 + \frac{6}{5} \frac{T_{vir}}{\bar{T}_{g} + \Delta T_{DM}}  \right) ^{3/2} -1
\end{equation}
where $\Delta T_{dm}$ is the extra heating due to dark matter annihilation. The $\Delta T_{dm}$ used here is the average change in temperature taken over the first 5 radial bins from the CGM results.

Figures 8 and 9 show the modified $\delta_{b}$ from DM annihilation for a $10^5\mathrm{M}_{\odot}$ halo at redshifts $20$ and $40$. Different colours correspond to different DM annihilating channels with blue, red, green, magenta and black showing the electron/positron, muon, quark, tau and W boson channels respectively. Different symbols as indicated in the bottom left hand plot indicate different DM masses. From left to right, the top plots show the fiducial and high halo models and the lower show results for Burkert and Burkert-high models.

\begin{figure}
\centering
\includegraphics[scale=0.45]{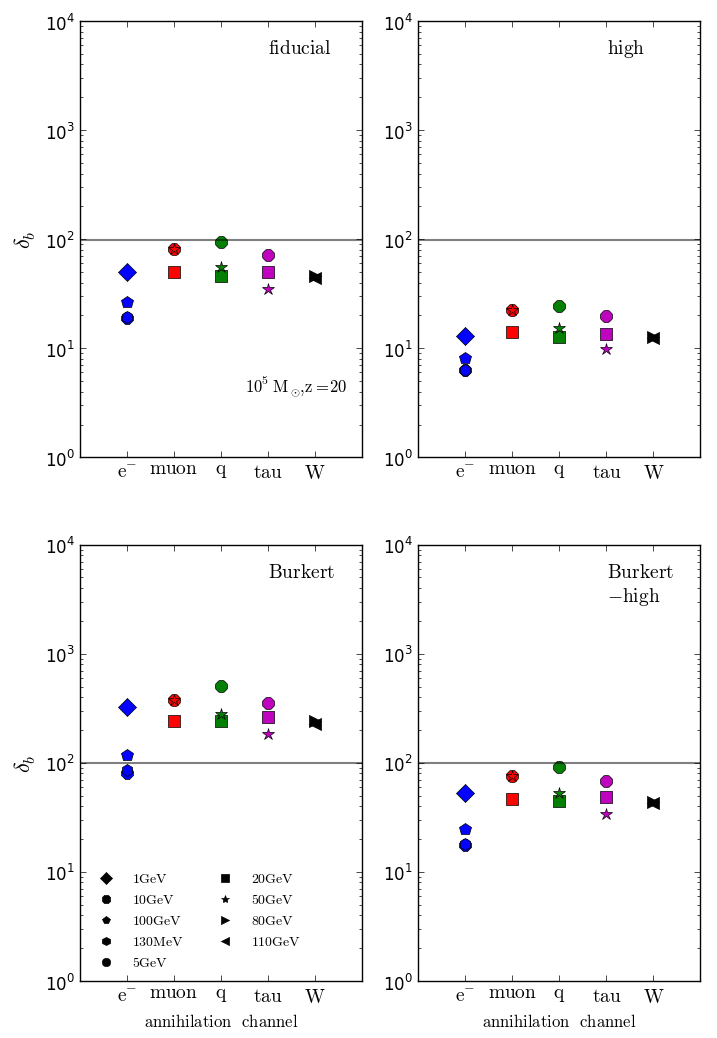}
\caption{Modification of $\delta_{b}$ from DM annihilation for a $10^5\mathrm{M}_{\odot}$ halo at redshift $20$. Different colours correspond to different DM annihilating channels with blue, red, green, magenta and black showing the electron/positron, muon, quark, tau and W boson channels respectively. Different symbols as indicated in the bottom left hand plot indicate different DM masses. From left to right, the top plots show the fiducial and high halo models and the lower show results for Burkert and Burkert-high models. The grey horizontal line shows the critical $\delta_{b}$ chosen to indicate the minimal $\delta_{b}$ required for baryonic collapse.}
\end{figure}

\begin{figure}
\centering
\includegraphics[scale=0.45]{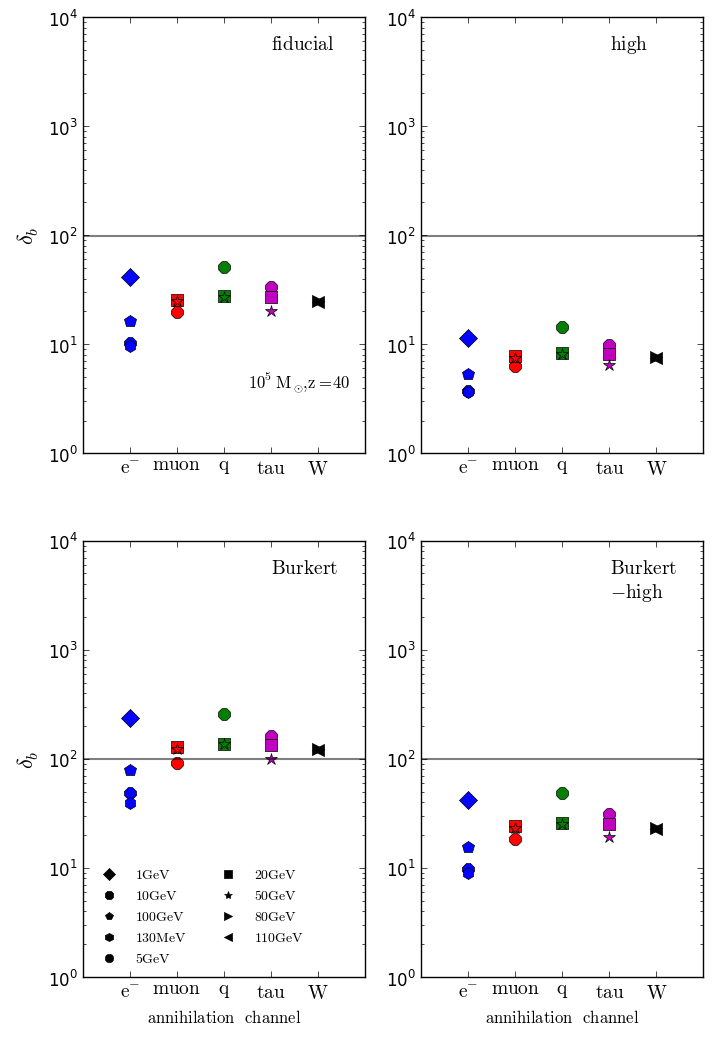}
\caption{Same as Figure 8 but for a $10^5\mathrm{M}_{\odot}$ halo at redshift $40$.}
\end{figure}

The results for the $10^5\mathrm{M}_{\odot}$ haloes in Figures 8 and 9 suggest that for a number of dark matter particle and halo models the additional heating from dark matter annihilation will lower $\delta_{b}$ below $100$. As expected this effect is most pronounced for the highly concentrated Einsto-High halo model due to the increase in annihilation power. In comparison results for the standard Burkert model show the impact to be reduced by about an order of magnitude. Understandably the fiducial model falls between the two extremes with the concentrated Burkert profile again producing results comparable to that of the fiducial model. 

In terms of dark matter particles, the $130$ MeV model annihilating to an electron/positron pair is the most effective in reducing $\delta_{b}$ due to both its efficiency in depositing energy into the gas locally and the fact that none of the annihilation energy is siphoned into neutrinos. Amongst the remaining annihilation channels heavier candidates provide a greater change in $\delta_{b}$ compared to models with masses from $1-5$ GeV. This is in keeping with the previous discussion which showed that electrons and positrons within that energy range produce IC photons which tend to free stream at IGM densities. Results for the halo at redshift $40$ are qualitatively similar to those at redshift $20$ but with $\delta_{b}$ reduced even further.

While not shown here, results indicate that $\delta_{b}$ is not significantly impacted for the $10^6$ and $10^7\mathrm{M}_{\odot}$ haloes, though again the results are comparable to the lower mass cases in their general behaviour. However as was briefly alluded to earlier, the method of calculating $\delta_{b}$ does not take into account the virialization shocks surrounding the haloes and for $10^{7}\mathrm{M}_{\odot}$ haloes shock heating of the surrounding gas becomes significant. The derivation of $\delta_{mod}$ in Equation 16 however assumes the gas temperature remains at the temperature of the IGM and therefore the method does not produce reliable results for haloes of this mass. 

\begin{figure}
\centering
\includegraphics[scale=0.45]{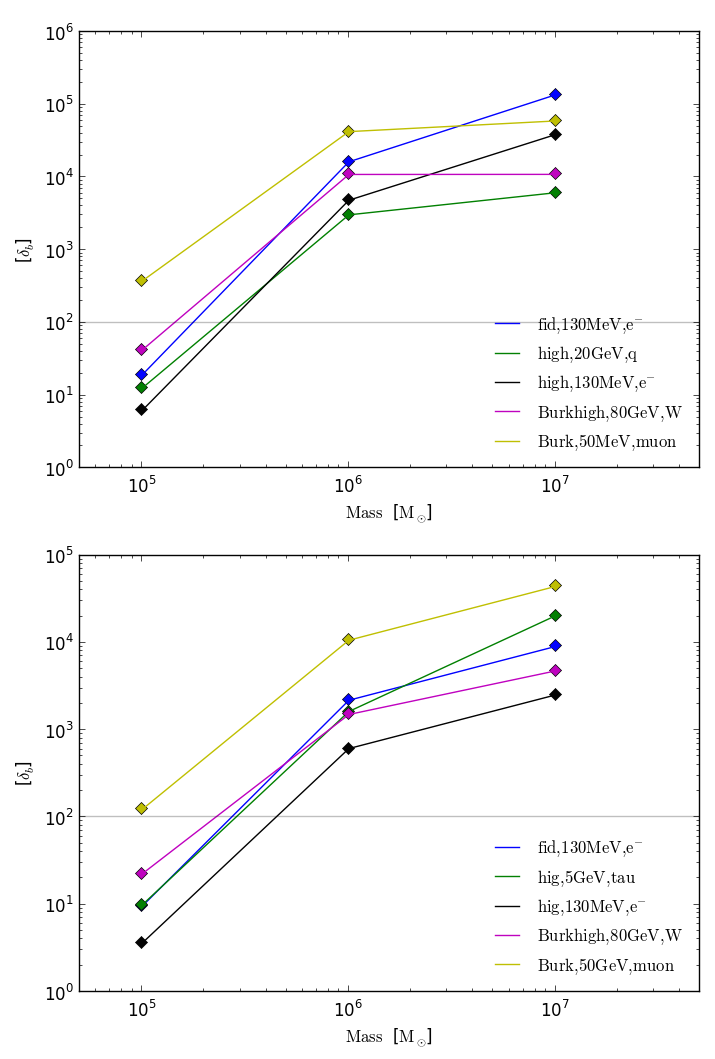}
\caption{Modification of $\delta_{b}$ from DM annihilation for different halo masses and dark matter models. The grey horizontal line again shows the critical $\delta_{b}$ chosen to indicate the minimum $\delta_{b}$ for sufficient gas accretion onto the halo.}
\end{figure}

\subsection{Change in the Minimal Baryonic Mass}
If $\delta_{b}$ is sufficiently reduced due to dark matter annihilation in $10^5\mathrm{M}_{\odot}$ haloes at redshifts $20$ and $40$, what consequences does this suggest for the minimal baryonic objects? Figure 10 again shows the annihilation-modified $\delta_{b}$ though this time as a function of halo mass. The top panel shows the results at redshift $20$ and the lower panel at redshift $40$. In both plots the grey, horizontal line denotes $\delta_{b}=100$ which gives the chosen critical value for the collapse of baryonic structures. Different coloured curves show the outputs for different halo/dark matter models.

For $\delta_{b}=100$ as the critical value, a number of dark matter candidates would reduce the infall of gas onto the halo to a degree that potential baryonic structure formation could be affected. For the dark matter model with the most pronounced impact on $\delta_{b}$, $130$ a MeV particle with a concentrated Einasto profile (black curve), a $10^5\mathrm{M}_{\odot}$ halo would have to increase in mass by a factor of $2-3$ at redshift $20$ and $4-5$ at redshift $40$ to recover a $\delta_{b}$ of $100$. It is again important to note that the large $\delta_{b}$ produced for haloes with mass  $>10^{6}\mathrm{M}_{\odot}$ arise due to the non-inclusion of virialization shocks and the subsequent shock heating of the gas surrounding the haloes. The derivation of $\delta_{mod}$ in Equation 21 however assumes the CGM gas temperature remains at the temperature of the IGM (which gives rise to the large $\delta_{b}$ in Figure 10) and therefore the method does not produce reliable results for haloes of this mass. This however does not impact the conclusions drawn for the $10^{5}\mathrm{M}_{\odot}$ haloes.

\section{Discussion}

In this paper we revisited the energy deposition from dark matter annihilation in small, high redshift haloes, as well as calculating the heating of the haloes' circumgalactic medium due to dark matter annihilation in the halo proper. A key aspect of this work is the updated energy transfer Monte Carlo code which allows for injected electrons, positrons and photons to interact with an arbitrary, three dimensional number density distribution of atomic hydrogen, helium, free electrons and CMB photons. The code calculates the fraction and location of the energy deposited into heating, ionisation and Lyman photons in a pre-defined volume such as a halo.

Overall the broad conclusions drawn in the earlier calculation are supported in the detailed study. We find that the total annihilation power is increased for more concentrated haloes, with the total annihilation power being significantly impacted by the precise distribution of dark matter within the halo. We also find the existence of a parameter space in which DM annihilation could modify baryonic structure formation. In our previous paper the choice of annihilation channel influenced outcomes predominantly through the fraction of the total annihilation power partitioned into electrons/positrons and photons, as opposed to the non-interacting neutrinos. Since the updated energy transfer method is sensitive to the energy dependent mean-free path, both the DM particle mass (for an annihilation cross section adjusted to give the same total power integrated over the halo) and choice of annihilation channel also impact on the overall energy deposition by changing the energy distribution of annihilation products. The $130$ MeV DM model annihilating via electron/positron pairs was particularly efficient at depositing energy with haloes due to the favourable interaction mean-free paths of the up-scattered IC photons produced by the injected particles.

We built on our detailed examination of self-heating haloes by calculating the heating of the haloes' circumgalactic medium. Sufficient increase in temperature of the CGM can lead to the suppression of gas infall onto the halo and formation of baryonic structure. $\delta_{mod}$ was calculated for the different DM models. At redshifts $20$ and $40$, a $10^5\mathrm{M}_{\odot}$ halo is no longer massive enough for $\delta_{b} > 100$, instead objects more massive by a factor of 3-5 are required. An important caveat is that the method used in the work breaks down for more massive haloes as it does not take into account shock heating haloeswhich results in an inflated value of $\delta_{b}$.

The work completed here aims to contribute to the identification of the parameter space in which dark matter annihilation impacts early structure formation as well as developing tools to assist with the rigorous treatment of energy transfer in non- homogeneous density fields. While the analysis of a single halo cannot predict the accumulative effect dark matter annihilation has on hierarchical structure formation over time, it does aid in the formulation of models that will be compatible with simulations and allow for testable predictions to be made.

In order to identify a signal indicative of new physics, careful modelling is required to account for both uncertainties in the dark matter and astrophysical models. It is crucial that predictions whether they be of a dark matter modified 21cm signal or general galaxy properties, integrates dark matter annihilation into the emergence of structure starting at high redshifts. A number of challenges remain before the latter can be fully realised but the prospective comparisons to be made with future observations will provide valuable constraints to guide the ongoing quest to uncover the fundamental nature of dark matter.

\bibliographystyle{mn2e}	
\bibliography{paper_bibliography}


\end{document}